# Cross sections from proton irradiation of thorium at 800 MeV

Jonathan W. Engle, Stepan G. Mashnik, John W. Weidner, Laura E. Wolfsberg, Michael E. Fassbender, Kevin Jackman, Aaron Couture, Leo J. Bitteker, John L. Ullmann, Mark S. Gulley, Chandra Pillai, Kevin D. John, Eva R. Birnbaum, and Francois M. Nortier

*Los Alamos National Laboratory, Los Alamos*
*P.O. Box 1663, Los Alamos, New Mexico, 87545*

*e-mail address: jwengle@lanl.gov*

**Abstract:** Nuclear formation cross sections are reported for 65 nuclides produced from 800-MeV proton irradiation of thorium foils. These data are useful as benchmarks for computational predictions in the ongoing process of theoretical code development and also to the design of spallation-based radioisotope production currently being considered for multiple radiotherapeutic pharmaceutical agents. Measured data are compared with the predictions of three MCNP6 event generators and used to evaluate the potential for 800-MeV productions of radioisotopes of interest for medical radiotherapy. In only a few instances code predictions are discrepant from measured values by more than a factor of two, demonstrating satisfactory predictive power across a large mass range. Similarly, agreement between measurements presented here and those previously reported is good, lending credibility to predictions of target yields and radioimpurities for high-energy accelerator-produced radionuclides.

## I. INTRODUCTION

High-energy transport codes are often used to estimate residual nuclide quantities produced from complex irradiation schemes. When nuclear formation cross sections for radioisotopes of interest have not been measured, the codes permit consideration of a planned irradiation's radioisotopic product distribution and quantity, as well as its economic and safety-related consequences. Measured data are essential to the development and validation of these codes. One such transport code, MCNP6, is employed by a broad user base for a wide variety of tasks [1]. Its efficacy is directly dependent upon the quality of its "event generators", which implement various models incorporated in Monte Carlo modules to simulate the interactions of individual particles with targets of specified geometries.

The 800-MeV proton beam at the Los Alamos Neutron Science Center (LANSCE) has been used to make a wide variety of radionuclides since the early 1970s [2]. The facility has recently been targeted for significant upgrades, which would re-establish milliampere-scale, spallation-based production of a variety of radioisotopes operating parasitically with world-leading materials testing and neutron-scattering capabilities [3]. Recent experiments used a combination of code predictions and activation of thin $^{232}$Th foils at the LANSCE facility to assess the potential for 800-MeV accelerator production of $^{225}$Ac ($t_{1/2}$ = 9.92 d) and $^{223}$Ra ($t_{1/2}$ = 11.4 d) for medical radiotherapeutic use [4, 5]. If they can be produced in sufficient quantity and radioisotopic purity, α-emitting radionuclides like $^{225}$Ac may one day be routinely bound with biological targeting vectors and introduced into human subjects as cell-killing agents with high selectivity for malignant tissue. In the wake of proof-of-principle studies, our attention has turned to the utility of the hypothetical product, a measure largely established by achievable radioisotopic purity and contextualized by available radiochemical separation techniques. Decay emissions of radioisotopic impurities often negate the benefits offered by the radioisotope of choice. Co-production of radioisotopic impurities such as $^{226}$Ac ($t_{1/2}$ = 29.37 h), which is chemically inseparable from the desired $^{225}$Ac, can only be avoided by careful control of irradiation parameters as well as the timing of chemical separation timing post irradiation. Radioisotopes of lower lanthanide elements are also expected to be challenging to remove from actinium solutions by established chromatographic methods (e.g. $^{139}$Ce and $^{141}$Ce, as well as $^{140}$La which is fed by production and decay of its non-lanthanide parent $^{140}$Ba).

Fig. 1 below illustrates some measure of the complexity of decay chains that must be considered by the code or by the researcher attempting production of a particular radioisotope of interest. Continuing with the example case of $^{225}$Ac, each isotope present in the decay chains of $^{225}$Ra, $^{226}$Ac, and $^{227}$Ac, as well as many others besides, is formed through production of its parents and through direct reactions with 800-MeV protons incident on targets of $^{232}$Th. A comprehensive understanding of the relevant physical mechanisms which form these nuclides, fed by increasing quantities of measured data and modeling by codes such those within MCNP6, will therefore serve to inform chemical separation investigations and ultimately the production methods necessary to make these and other radioisotopes available in useful forms.

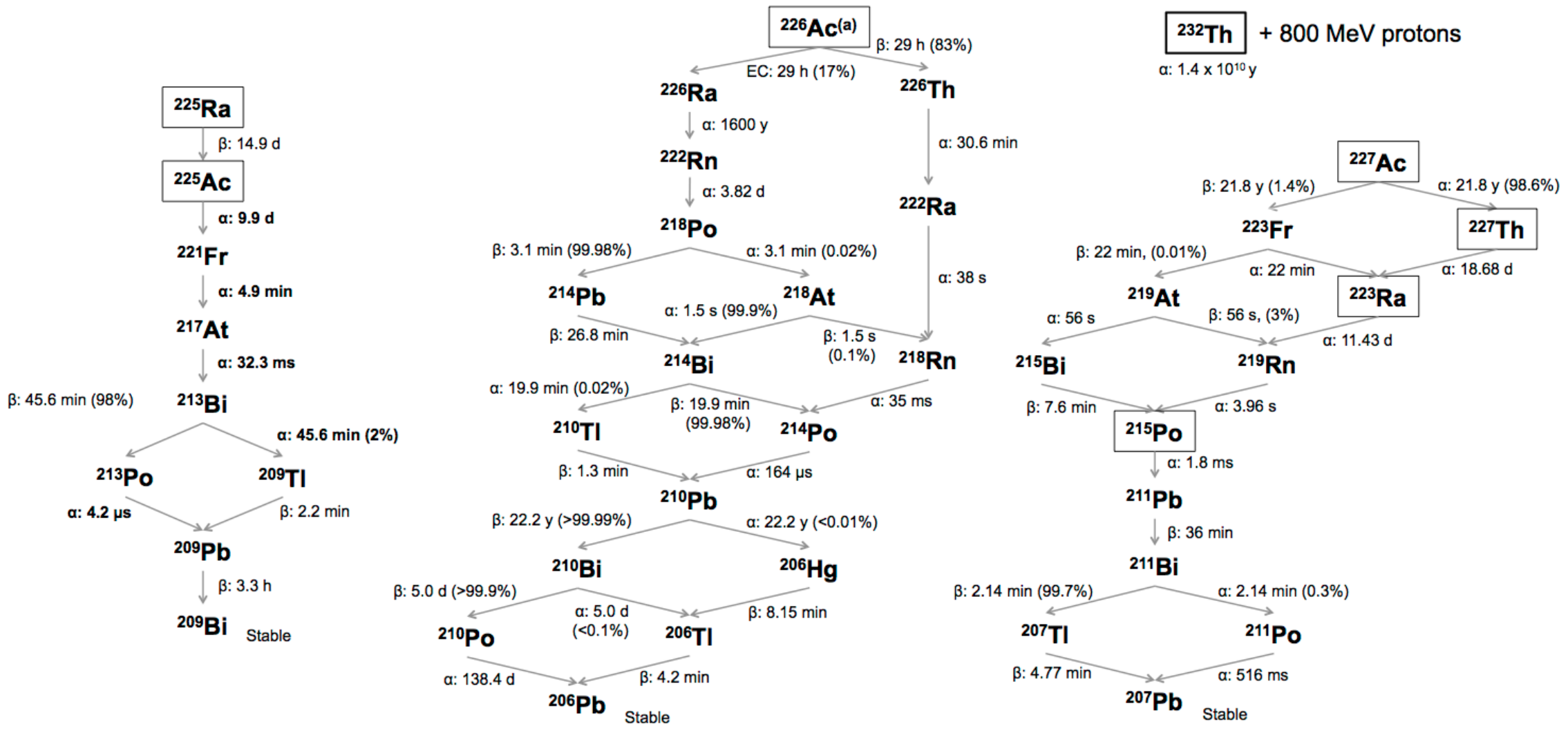

FIG. 1. The decay schemes for $^{225}$Ra, $^{226}$Ac, and $^{227}$Ac. Nuclear formation cross sections for isotopes enclosed in boxes are previously measured or reported in the present work. [(a)]The full decay scheme of $^{226}$Ac, starting with α-decay of $^{230}$Pa and including $^{226}$Ac's own α-decay branch (6 x $10^{-3}$ %) is not shown for simplicity.

## II. MATERIALS AND METHODS

The event generators of MCNP6 used in this work calculate only the direct formation of individual radionuclides. Predictions of produced quantities are therefore also integrally dependent upon the accuracy of nuclear data repositories to account for the contributions from "indirect reactions" induced by secondary particles of lower energy produced during the initial "direct reactions" and parent isotopes' decay. The two methods of isotope formation, by the decay of parents and by direct reaction, are described with the terms "cumulative" and "independent", respectively [6] and designated using the labels "c" and "i". When parents' decay is incompletely accounted for in the data, cross sections are described with the label "c*". Obviously, most yields of isotopes quantified by the activation technique will be cumulative, as their parents have long since decayed at the time of assay, preventing spectroscopic deconvolution of the signals from different isotopes within a single decay chain. In such cases, experiments utilizing inverse kinematics offer additional sensitivity, detecting residual nuclides formed by heavy ion bombardment of liquid hydrogen ($^1$H) within hundreds of nanoseconds following the reaction (see e.g. [7, 8] and references therein). The inverse kinematic method does not distinguish between ground and metastable states (designated "m" and "g" below) because of its short measurement timescale, whereas the activation method enables detection of especially longer-lived isomers when their decay emissions are quantifiable.

Generally, hundreds of radioisotopes are produced by spallation-, fission-, and evaporation-type reactions on thorium and other high-Z targets from incident particle energies in the hundreds of MeV. Individual reactions typically occur in two main stages. In the first, the so-called intranuclear cascade, incident particles interact with individual nucleons, instead of the nucleus as a whole. High-energy particles can leave the nucleus and potentially initiate further spallation reactions in neighboring nuclei, resulting in a chain reaction process whose intensity declines with the energy of secondarily emitted particles. In the second stage, the residual nucleus, now in an excited state, relieves its excitation energy through competing processes of evaporation and fission. If the excitation energy of the residual nucleus produced after the intranuclear cascade stage of a reaction is of the order of tens of MeV or greater, preequilibrium emission of particles is also possible during the equilibration of the nucleus, before final evaporation of particles or fission of the compound nucleus.

Measured data presented here are also compared with a significant body of literature on the proton irradiation of thorium targets not already mentioned above. Titarenko and coauthors presented an especially thorough investigation of formation cross sections to the International Atomic Energy Agency in 2003 from proton irradiation of numerous targets at energies between 100 and 2600 MeV; 22 of the cross sections reported here were not measured by this previous work [9]. Fission and fragmentation cross sections [10], fission fragment energies and angular distributions [20], absolute (p,xn)-type cross sections [11], alpha emissions relevant to radioactinide production [12], and pion production [13] have been previously investigated and reported in the energy range between 0.5 and 1 GeV. Additional work targeting specific radionuclide formation cross sections has reported data for $^{32}$P and $^{33}$P from 600 MeV protons [14], $^{83,84,86}$Rb [15], many Xe and Kr radioisotopes [16], and radionuclides from several actinide elements (see e.g. [17-20]) at energies hundreds of MeV above and below that used in this work. This work presents data for 65 radionuclides, several of which have not previously been measured, and is compared directly with the published values reported by Titarenko and coauthors [9] when applicable.

### A. Irradiation and gamma spectroscopy

A full description of the irradiation parameters used in this experiment has been reported previously [4]. Briefly, thin thorium foils were irradiated in the Target 2 Blue Room of the Weapons Neutron Research Facility at LANSCE for approximately one hour with 800-MeV protons at an intensity of 90 nA. Aluminum foils were irradiated simultaneously in order to use the $^{27}$Al(p,x)$^{22}$Na reaction as a monitor of integrated beam current. Stainless steel foils were also irradiated and exposed to Gafchromic film in order to confirm the beam's incidence on thorium and aluminum targets. So-called "blank" foil holders without sample foils were also simultaneously irradiated to permit subtraction of signals from isotopes produced secondary or scattered particle activation of non-target materials in the stack (e.g., for quantification of $^{7}$Be). Irradiated samples were transported to the Nuclear and Radiochemistry Group (C-NR) Countroom, where they were repeatedly assayed by non-destructive gamma spectroscopy for approximately 10 months. The HPGe detector used to assay the foils is a p-type aluminum windowed ORTEC GEM detector with a relative efficiency at 1333 keV of about 10% and a measured gamma peak FWHM at 1333 keV of 1.99 keV. Contributions to spectra backgrounds, detector resolution, and energy calibration (gain), were checked daily. Detector efficiency was calibrated prior to the beginning of data collection and verified after the experiment's completion. An extensively validated in-house analysis code, SPECANAL, was used to extract photopeak areas from gamma spectra for this work;

details of its methods are discussed elsewhere [21]. Gamma energies and intensities were taken from the National Nuclear Data Center's (NNDC) online archives [22]. The activity at the end of bombardment (EoB) of each isotope of interest was determined by fitting of its decay curve, and cross sections were calculated using the well-known activation formula.

Uncertainties in linear regressions' fitted parameters were computed from covariance matrices as the standard error in the activity extrapolated to the end of bombardment. This value was combined according to the Gaussian law of error propagation with estimated contributing uncertainties from detector calibration and geometry reproducibility (2.9% combined), target foil dimensions (0.1%), and proton flux (6.7%). Multiple photopeaks were used (up to a maximum of 4) when possible, and so additional uncertainty as the standard deviation of these complimentary measurements was combined with the uncertainties described above, again according to the Gaussian law of error propagation.

## B. MCNP6 event generators tested here

We compare our measured cross sections with predictions of the MCNP6 transport code [1, 23] using three different event generators available in MCNP6 to simulate high energy nuclear reactions. All predictions were obtained prior to the measurement. These event generators have previously been benchmarked against a large variety of experimental data and compared with each other and several other modern models (see e.g. [6] and references therein).

A brief description of the three event generators follows:

1) The default MCNP6 option, which for our reaction is an improved version of the Cascade-Exciton Model (CEM) of nuclear reactions as implemented in the code CEM03.03 [24, 25].

2) The Bertini IntraNuclear Cascade (INC) [26], followed by the Multistage Preequilibrium Model (MPM) [27], followed by the evaporation model as described with the EVAP code by Dresner [28], followed by or in competition with the RAL fission model [29] (if the charge of the compound nucleus Z is ≥ 70), referred to herein simply as "Bertini".

3) The IntraNuclear Cascade model developed at the Liege (INCL) University in Belgium by Prof. Cugnon with his coauthors from CEA, Saclay, France [30] merged with the evaporation-fission model ABLA [31] developed at GSI, Darmstadt, Germany, referred to herein as "INCL+ABLA".

The improved Cascade-Exciton Model (CEM) as implemented in the code CEM03.03 [24, 25] calculates nuclear reactions induced by nucleons, pions, and photons. It assumes that the reactions occur generally in three stages: The first stage is the INC, in which primary particles can be re-scattered and produce secondary particles several times prior to absorption by (or escape from) the nucleus. When the cascade stage of a reaction is completed, CEM03.03 uses the coalescence model to "create" high-energy d, t, $^{3}He$, and $^{4}He$ particles by final-state interactions among emitted cascade nucleons. The emission of the cascade particles determines the particle–hole configuration, Z, A, and the excitation energy that is the starting point for the second, preequilibrium stage of the reaction. The subsequent relaxation of the nuclear excitation is treated with an improved version of the modified exciton model of preequilibrium decay followed by the equilibrium evaporation/fission stage (also called the compound nucleus stage), which is described with an extension of the Generalized Evaporation Models (GEM) code, GEM2, by Furihata [32]. Generally, all components may contribute to experimentally measurable particle emission spectra and affect the final residual nuclei. But if the residual nuclei after the INC have atomic numbers in the range A < 13, CEM03.03 uses the Fermi breakup model [33] to calculate their further disintegration instead of using the preequilibrium and evaporation models. Fermi breakup is faster to calculate than GEM and gives results similar to the continuation of the more detailed models to much lighter nuclei.

In MCNP6, by default, Bertini INC [26] is followed by the Multistage Preequilibrium Model (MPM) [27]. The relaxation of an excited compound nucleus produced after the preequilibrium stage of a reaction is calculated with the Weisskopf evaporation model as implemented in the EVAP code by Dresner [28]. If the charge of the compound nucleus Z < 70, then a competition between evaporation and fission is taken into account, with the latter calculated using the RAL fission model by Atchison [29]. The Bertini default option of MCNP6 also accounts for Fermi breakup of excited nuclei when A < 18, but does not account for the coalescence of complex particles from INC nucleons.

The version of INCL [30] available at present in MCNP6 is usually used to describe reactions induced by nucleons and complex particles up to $^{4}$He at incident energies up to several GeV. In MCNP6, it is merged with the evaporation–fission model ABLA [31] developed at GSI in Darmstadt, Germany. The version of INCL + ABLA available currently in MCNP6 accounts for possible fission of compound nuclei produced in our reaction, but it does not account for preequilibrium processes, for Fermi break-up of light residual nuclei, or for coalescence of complex particles after (or during) INC.

All event generators used compute only independent cross sections; cumulative cross sections were subsequently calculated using these independent values summed separately according to the decay behavior of parent products using the Table of Isotopes [34].

## III. RESULTS AND DISCUSSION

### A. Cross sections

Tabulated results are presented below. In a few isolated cases the uncertainties on reported cross sections are large; in these cases poor counting statistics or especially challenging peak fitting directly contributed to the high reported uncertainty. Nevertheless, general agreement between measured and calculated cross sections across the mass range of cumulative cross sections measured is generally good (Fig. 2). As has been reported previously for reactions on target nuclei with lower masses, MCNP6 event generators do poorly accounting for production of very low-Z nuclei, such as $^{7}$Be (Fig. 3a), which are likely produced in the evaporative stages of compound nuclei relaxation following spallation events [21], as well as in other fragmentation processes, like multifragmentation and/or fission-like binary decays, not accounted yet by the current version of MCNP6. Agreement between event generator predictions and measured data is acceptable across the range of fission products (Fig. 3b).

MCNP6 event generators are in uniform disagreement with measured data for the cumulative cross sections of $^{203}$Bi and $^{205}$Bi (Fig. 3c). Presently we have neither a clear explanation for this nor for several other significant disagreements between the calculated values and our measured data. A further, more detailed, investigation is needed. The comparisons clearly demonstrate that all models tested here must be improved in order to accurately predict yields of isotopes from arbitrary reactions.

TABLE I. A comparison of measured data with previously measured values from [9] and with calculated cross sections from CEM03.03, Bertini, and INCL + ABLA event generators of MCNP6 for 800-MeV proton irradiation of thorium foils.

| Isotope | $t_{1/2}$ (d) | Type[a] | Measured cross sections | | | | MCNP6 calculated cross sections | | |
|---|---|---|---|---|---|---|---|---|---|
| | | | Current work | | Ref. [9] | | CEM03.03[b] | Bertini[b] | INCL+ABLA[b] |
| | | | σ (mb) | Δσ (mb) | σ (mb) | Δσ (mb) | σ (mb) | σ (mb) | σ (mb) |
| $^{7}$Be | 53.29 | i | 1.3 | 0.6 | - | - | 0.37 | 0.00 | 0.00 |
| $^{46}$Sc | 83.79 | i(m+g) | 0.7 | 0.1 | - | - | 0.10 | 0.33 | 0.13 |
| $^{48}$Sc | 1.82 | i | 0.4 | 0.2 | 0.9 | 0.2 | 0.02 | 0.40 | 0.07 |
| $^{59}$Fe | 44.50 | c | 2.0 | 0.3 | - | - | 1.31 | 2.74 | 1.32 |
| $^{74}$As | 17.77 | i | 4.4 | 0.3 | - | - | 3.87 | 2.14 | 1.81 |
| $^{76}$As | 1.09 | i | 4.4 | 0.3 | 5.0 | 0.4 | 4.24 | 2.93 | 4.10 |
| $^{77}$Br | 2.38 | c | 0.9 | 0.2 | - | - | 3.83 | 2.61 | 0.89 |
| $^{82}$Br | 1.47 | i(m+g) | 8.5 | 0.9 | 8.6 | 0.7 | 5.95 | 4.39 | 7.35 |
| $^{86}$Rb | 18.64 | i(m+g) | 15.6 | 1.2 | - | - | 10.10 | 8.10 | 6.36 |
| $^{87}$Y | 3.33 | c | 3.7 | 0.7 | 3.5 | 0.2 | 7.44 | 4.64 | 1.93 |
| $^{87m}$Y | 0.56 | c | 4.1 | 0.4 | 2.8 | 0.5 | 7.44 | 4.64 | 1.93 |
| $^{88}$Y | 106.63 | i | 6.4 | 0.4 | - | - | 8.13 | 4.43 | 3.34 |
| $^{91}$Sr | 0.40 | c | 18.9 | 1.6 | 22.8 | 2.1 | 8.52 | 13.23 | 25.74 |
| $^{95}$Nb | 34.99 | i(m+g) | 14.2 | 0.9 | 12.7 | 1.3 | 14.50 | 6.41 | 9.89 |
| $^{95}$Tc | 0.83 | c* | 1.2 | 0.3 | 1.0 | 0.1 | 3.55 | 2.42 | 0.61 |
| $^{95}$Zr | 64.03 | c | 30.8 | 2.0 | 31.5 | 2.9 | 37.72 | 19.47 | 29.03 |
| $^{96}$Nb | 0.97 | i | 15.3 | 1.2 | 14.8 | 1.0 | 12.80 | 5.86 | 13.33 |
| $^{96}$Tc | 4.28 | i(m+g) | 2.1 | 0.2 | 3.0 | 0.9 | 3.97 | 2.27 | 1.58 |
| $^{99}$Mo | 2.75 | c | 39.1 | 2.6 | 45.0 | 3.1 | 20.66 | 19.15 | 39.45 |
| $^{99m}$Tc | 0.25 | i(m) | 2.0 | 0.4 | 2.5 | 0.2 | 14.13 | 8.22 | 9.39 |
| $^{100}$Pd | 3.7 | c | 1.1 | 1.2 | - | - | 0.76 | 0.72 | 0.08 |
| $^{100}$Rh | 0.87 | c | 5.0 | 2.9 | - | - | 5.02 | 2.81 | 0.87 |
| $^{103}$Ru | 39.26 | c | 52.1 | 4.3 | 61.0 | 4.4 | 28.23 | 21.14 | 44.49 |
| $^{105}$Rh | 1.47 | c | 55.2 | 3.4 | 52.1 | 3.9 | 33.36 | 27.05 | 45.85 |
| $^{105}$Ru | 0.19 | c | 37.1 | 3.1 | 43.0 | 2.9 | 17.56 | 18.11 | 34.36 |
| $^{106m}$Ag | 8.28 | i(m) | 2.4 | 0.4 | - | - | 3.98 | 2.78 | 1.99 |
| $^{110m}$Ag | 249.76 | i(m) | 11.3 | 1.1 | - | - | 9.17 | 5.13 | 10.48 |
| $^{111}$In | 2.80 | c | 3.0 | 0.3 | 3.0 | 0.3 | 5.88 | 3.90 | 2.25 |
| $^{114m}$In | 49.51 | i(m) | 9.0 | 0.6 | - | - | - | - | - |
| $^{115}$Cd | 2.27 | c | 21.5 | 2.3 | 21.1 | 1.6 | 14.63 | 15.61 | 27.97 |
| $^{117m}Sn$ | 13.76 | i(m) | 8.9 | 0.6 | 7.0 | 1.1 | 6.20 | 4.14 | 6.96 |
| $^{120m}Sb$ | 5.76 | i(m) | 6.7 | 0.5 | 7.0 | 0.5 | 5.18 | 3.87 | 6.63 |
| $^{121}Te$ | 19.17 | c | 3.5 | 0.3 | - | - | 8.61 | 6.22 | 6.09 |
| $^{121m}Te$ | 164.20 | i(m) | 13.4 | 10.5 | 5.3 | 0.5 | 4.23 | 6.22 | 6.09 |
| $^{122}Sb$ | 2.73 | i(m+g) | 8.2 | 0.5 | 8.7 | 0.6 | 4.96 | 4.46 | 4.46 |
| $^{123}I$ | 0.55 | c | 7.5 | 0.8 | - | - | 6.71 | 5.66 | 4.57 |
| $^{123m}Te$ | 119.20 | i(m) | 6.0 | 0.4 | - | - | 4.53 | 3.99 | 6.14 |
| $^{124}I$ | 4.17 | i | 5.4 | 1.5 | 4.7 | 0.7 | 4.33 | 3.86 | 4.14 |
| $^{124}Sb$ | 60.20 | i(m1+m2+g) | 7.3 | 0.5 | 8.5 | 2.4 | 3.59 | 3.51 | 6.54 |
| $^{126}I$ | 12.93 | i | 6.1 | 0.5 | - | - | 4.14 | 4.30 | 5.27 |
| $^{126}Sb$ | 12.46 | i(m1+m2+g) | 4.0 | 0.7 | 3.2 | 0.3 | 1.78 | 2.87 | 5.29 |
| $^{127}Sb$ | 0.16 | c | 3.6 | 0.3 | 5.1 | 0.5 | 1.81 | 6.86 | 6.98 |
| $^{127}Xe$ | 36.35 | c | 5.6 | 2.8 | 8.4 | 1.0 | 5.30 | 4.76 | 5.94 |
| $^{130}I$ | 0.52 | i(m+g) | 4.3 | 1.7 | 4.1 | 0.3 | 2.40 | 3.32 | 4.89 |
| $^{131}Ba$ | 11.50 | c | 4.1 | 0.6 | - | - | 3.38 | 3.90 | 3.26 |
| $^{131}I$ | 8.03 | c | 6.1 | 0.4 | 6.8 | 0.5 | 3.23 | 8.57 | 11.62 |
| $^{133}I$ | 0.87 | c | 4.2 | 0.9 | 4.6 | 0.4 | 1.66 | 6.13 | 8.58 |
| $^{134}Cs$ | 754.31 | i(g) | 3.1 | 0.2 | - | - | 2.18 | 2.47 | 4.06 |
| $^{135}I$ | 0.28 | c | 2.0 | 0.2 | 2.9 | 0.3 | 0.53 | 3.71 | 4.85 |
| $^{135}Xe$ | 0.38 | c | 2.2 | 0.1 | 6.9 | 0.5 | 7.52 | 10.34 | 11.48 |
| $^{136}Cs$ | 13.16 | i(m+g) | 2.0 | 0.2 | - | - | 2.23 | 2.58 | 4.67 |
| $^{139}Ce$ | 137.64 | c | 4.1 | 0.3 | - | - | 2.71 | 5.22 | 4.64 |
| $^{140}Ba$ | 12.75 | c | 2.8 | 0.4 | 5.2 | 0.9 | 1.99 | 4.94 | 7.47 |
| $^{143}Ce$ | 1.38 | c | 3.8 | 0.3 | 4.2 | 0.3 | 2.69 | 4.43 | 6.69 |
| $^{188}Pt$ | 10.20 | c | 3.4 | 0.3 | - | - | 11.46 | 11.73 | 1.51 |
| $^{200}Pb$ | 0.90 | c | 6.3 | 1.2 | 7.7 | 0.6 | 3.09 | 9.28 | 3.84 |
| $^{203}Bi$ | 0.49 | c | 8.0 | 3.0 | 10.3 | 0.8 | 1.76 | 1.41 | 0.75 |
| $^{204}Po$ | 0.15 | c | 5.0 | 0.9 | 13.5 | 1.1 | 8.84 | 12.14 | 5.37 |
| $^{205}Bi$ | 15.31 | c | 12.8 | 1.1 | 18.2 | 1.7 | 1.11 | 0.77 | 0.59 |
| $^{206}Po$ | 8.80 | c | 21.1 | 2.4 | 20.0 | 1.5 | 13.59 | 12.16 | 7.77 |
| $^{207}Po$ | 0.24 | c* | 18.6 | 2.6 | - | - | 19.56 | 16.89 | 9.07 |
| $^{209}At$ | 0.23 | c* | 23.4 | 2.6 | 17.8 | 1.2 | 23.45 | 15.33 | 11.32 |
| $^{210}At$ | 0.34 | c | 10.3 | 2.1 | 11.0 | 0.8 | 14.69 | 8.60 | 8.47 |
| $^{211}Rn$ | 0.61 | c | 9.6 | 2.0 | 9.9 | 0.8 | 9.51 | 6.87 | 7.12 |
| $^{226}Ac$ | 1.22 | c | 13.3 | 3.1 | 16.6 | 1.6 | 13.60 | 10.54 | 13.73 |

[a] Refers to the type of cross section measured: c = cumulative; $c^*$ = cumulative, with only partial accounting for possible parent contributions through decay; i = independent, with isomeric states parenthetically identified.

[b] All models provide only the sum of isomeric states of the isotope, and therefore, calculated predictions may overestimate independent, measured cross sections of a single isomeric state.

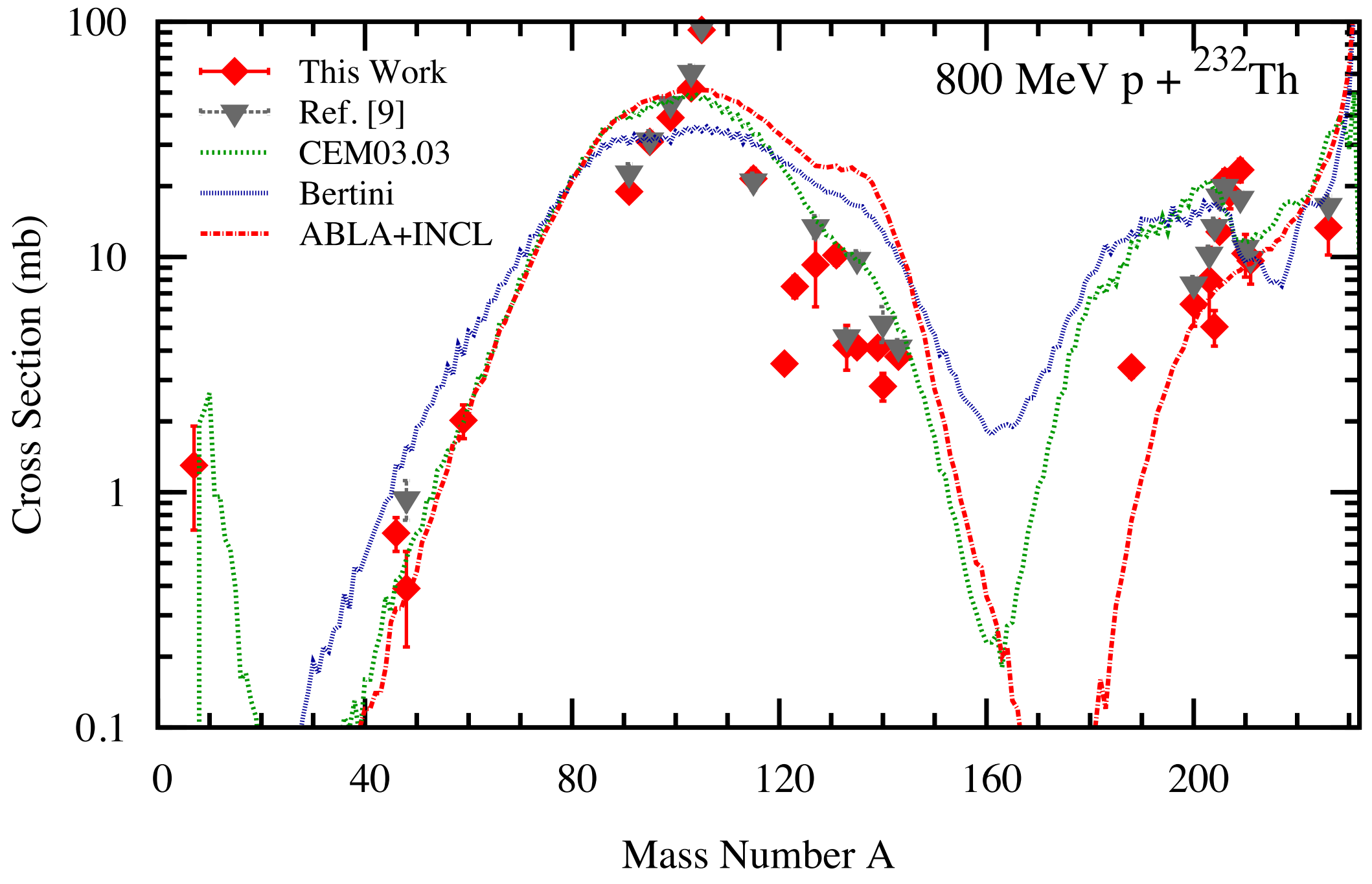

FIG. 2. (Color online) Comparison of mass distributions of product yields predicted by CEM03.03, Bertini, and INCL+ABLA from 800-MeV p + $^{232}$Th with cumulative cross sections measured in the present work and those measured previously by Titarenko and coauthors [9]. Where cumulative cross sections for isotopes of the same mass number are measured on both sides of the valley of stability, the cross sections are summed to accurately reflect the values calculated by MCNP6 event generators.

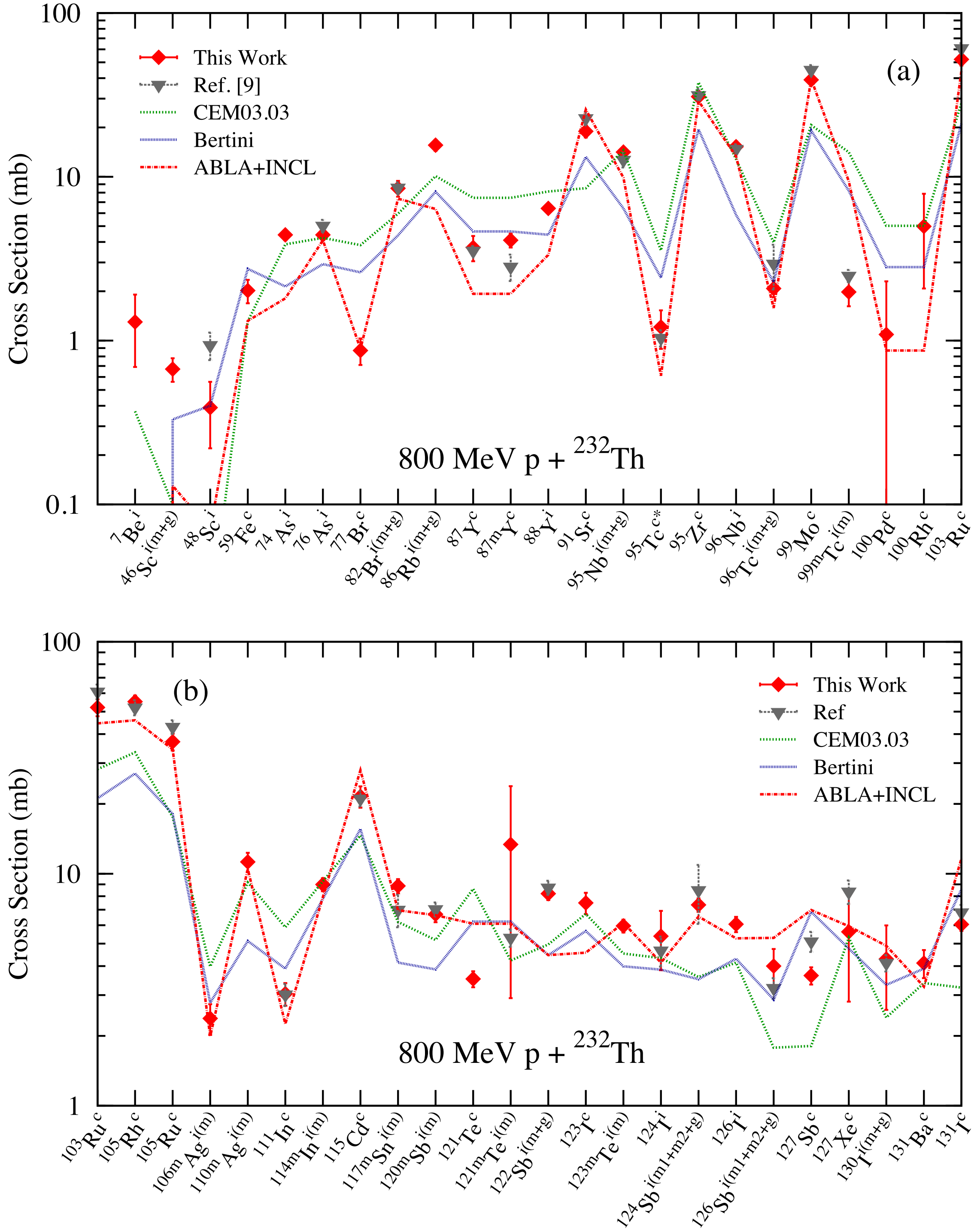

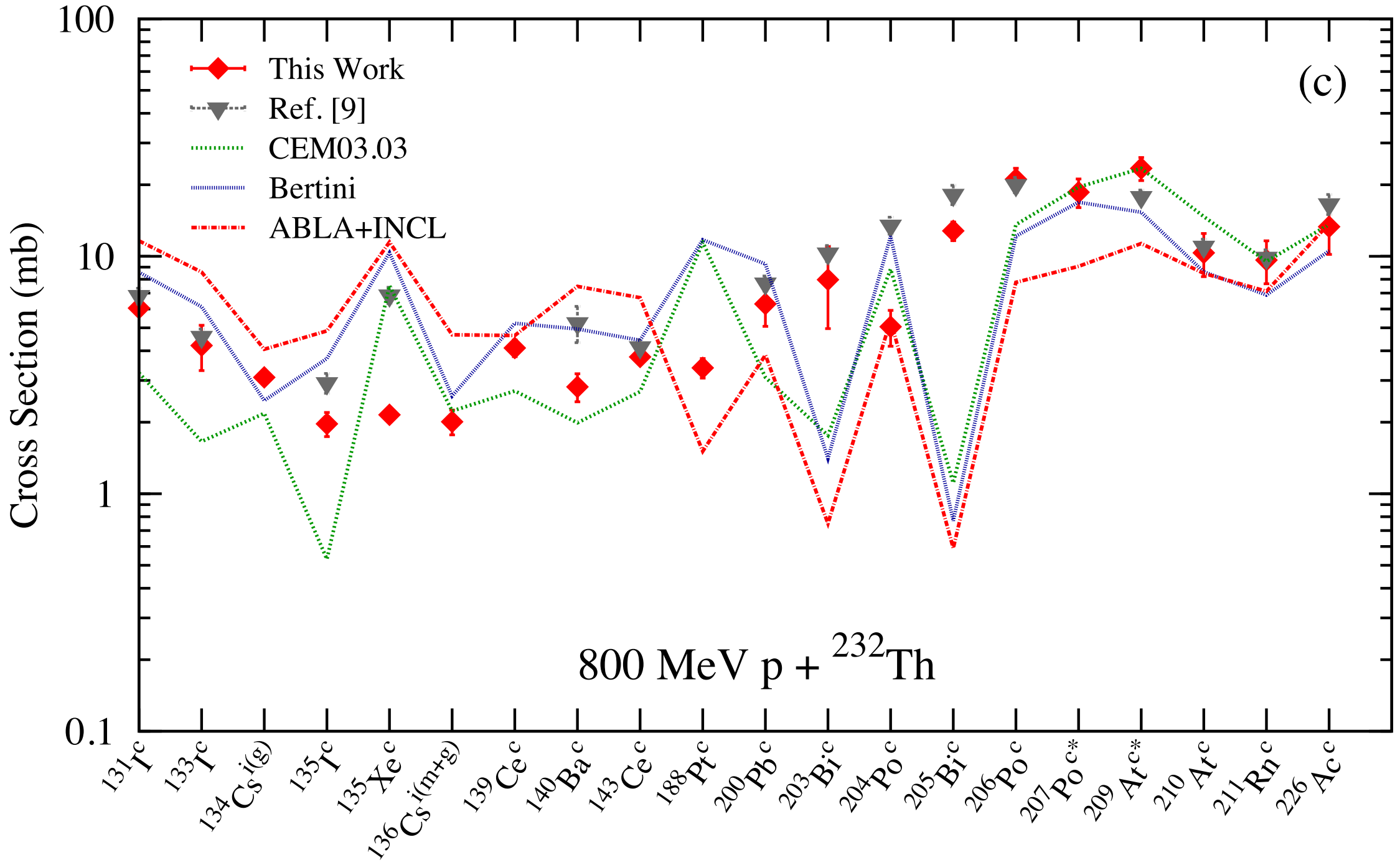

FIG. 3. (Color online) Detailed comparison between all cross sections measured in the present work, previously measured values from [18], and the predictions of CEM03.03, Bertini, and INCL+ABLA event generators of MCNP6 in order of increasing mass. Independent (i) and cumulative (c) cross sections are identified by superscripts following the isotope label on the x-axes.

### B. Predicted yields and radiochemical purity of isotopes of interest

Developmental efforts targeting production of tens of Curies of $^{225}$Ac and/or $^{223}$Ra for medical radiotherapeutic applications must contend with the challenge of radiochemically separating microgram quantities of desired actinides from gram-scale thorium targets and dozens of fission products. The numerous cross sections reported here and elsewhere in the literature speak to the difficulty of this task. Furthermore, radioisotopes of the same element as desired products can only be minimized by adjustment of irradiation parameters and by allowing final products to decay; chemical separation techniques will not be helpful in resolving this latter problem. For example, $^{226}$Ac and $^{227}$Ac will be co-produced during any irradiation targeting $^{225}$Ac. Their relative yields, combined with patients' tolerance for these radioactive impurities *in vivo*, will determine the timeframe within which a finished product containing $^{225}$Ac can be injected into the human subject and the maximum activity that such an injection may contain.

Any production-scale irradiation of a thorium target with 800-MeV protons represents a major undertaking, likely producing hundreds to thousands of curies of total activity at the end of days-long bombardments. For this reason, and because the intense effort and resource investment involved are compounded by the necessity for extensive chemical processing, invested parties will doubtless seek maximal return on their investment of time and resources. The most obvious way to increase this return is to simultaneously "harvest" multiple radioisotopes of interest from each individual irradiation. For this reason, Table 2 below details the predicted yields and instantaneous production rates using a selection of the cross sections measured here generalized to a 10 day, 1250 μA irradiation of a 5 g cm$^{-2}$ thorium target with 800-MeV protons. These irradiation parameters are representative of those that would be used at LANSCE for spallation-based production of radioisotopes in the "Area A" facility, which has used beams with similar intensities in the past. Isotopes are selected for their relevance to the production schemes of actinides of interest or because they are sources of general scientific interest.

TABLE II. A selection of yields (Bq and Ci) at the end of bombardment EoB and instantaneous production rates at $t_0$ from a representative 10 day, 1250 μA irradiation of a 5g cm$^{-2}$ Th target with 800-MeV protons.

| Nuclide | $t_{1/2}$ | EoB Yield in Bq (Ci) | Instantaneous Production Rate in MBq μA$^{-1}$ hr$^{-1}$ (μCi μA$^{-1}$ hr$^{-1}$) | Reason for Inclusion |
|---|---|---|---|---|
| $^{225}$Ac | 9.9 d | 7.47E+11 (20.2) | 3.46 (93.6) | Desired α-emitter [a] |
| $^{226}$Ac | 29.0 h | 1.36E+12 (36.6) | 4.52 (122.1) | Impurity in $^{225}$Ac production |
| $^{227}$Ac | 21.8 y | 1.84E+9 (0.05) | 0.006 (0.17) | Impurity in $^{225}$Ac production [a] |
| $^{225}$Ra | 14.8 d | 1.26E+11 (3.4) | 0.52 (14.0) | Potential $^{225}$Ac or $^{213}$Bi generator [a] |
| $^{223}$Ra | 11.4 d | 2.44E+11 (6.6) | 1.08 (29.3) | Impurity in $^{225}$Ra production [a] |
| $^{203}$Bi | 0.49 d | 8.09E+11 (21.9) | 2.70 (72.8) | Impurity in $^{225}$Ac/$^{213}$Bi generator system |
| $^{205}$Bi | 15.3 d | 3.56E+12 (96.2) | 11.89 (320.8) | Impurity in $^{225}$Ac/$^{213}$Bi generator system |
| $^{140}$Ba | 12.8 d | 6.81E+11 (18.4) | 2.27 (61.4) | Parent of $^{140}$La, which follows Ac chemistry |
| $^{139}$Ce | 137.6 d | 8.49E+12 (229.4) | 28.29 (764.7) | Follows Ac chemistry |
| $^{143}$Ce | 33.0 h | 3.84E+11 (10.4) | 1.28 (34.6) | Follows Ac chemistry |
| $^{99}$Mo | 66.0 h | 4.31E+12 (116.4) | 14.35 (388.0) | Nationally needed isotope for nuclear medicine |

[a] Reference [4].

## IV. CONCLUSIONS

Cross sections for 65 nuclides produced by the 800-MeV proton irradiation of thorium were measured and compared with existing data measured by Titarenko et al. at ITEP, Moscow [9] as well as with predictions by CEM03.03 [24, 25], Bertini+ MPM+Dresner+RAL [26-29], and INCL+ABLA [31, 31] event generators of the MCNP6 transport code [1]. Calculations by all event generators agree well with the measured data and with each other in the mass region above A ~ 200, where spallation reactions dominate. Calculated values are also in acceptable agreement with measured cross sections for fission fragment isotopes with mass numbers from A ~ 46 to A ~ 143. The quality of agreement between the codes and measured data for radioisotopes with Z > 80 is, with the exception of $^{203}$Bi and $^{205}$Bi, particularly encouraging in light of the codes' ability to accurately predict yields from production irradiations targeting $^{225}$Ac for medical radiotherapy.

The greatest disagreement between MCNP6 event generators' predictions and measured values exists in the transition regions between spallation and fission reactions and between fission and fragmentation reactions. In these regions, disagreement between calculated and measured data approaches an order of magnitude, and disagreement between individual event generators exceeds two orders of magnitude, for products with mass number near A = 180.

Only one product, $^{7}$Be, could be measured in the fragmentation region. CEM03.03 predicts a cross section for $^{7}$Be about 3.5 times lower than the measured value, while INCL+ABLA and Bertini do not predict any formation of $^{7}$Be products from 800-MeV protons incident on thorium. Further experimental data near the transition between spallation and fission reactions and between fission and fragmentation reactions would improve understanding of the mechanisms of these nuclear reactions and assist with further possible improvement of models. The computational models studied here are expected to benefit from modifications to improve their predictive accuracy in light of these measured data.

## V. ACKNOWLEDGEMENTS

We are grateful for technical assistance from LANL C-NR, C-IIAC, AOT-OPS, and LANSCE-NS groups' staff. This study was carried out under the auspices of the National Nuclear Security Administration of the U.S. Department of Energy at Los Alamos National Laboratory under Contract No. DE-AC52-06NA253996 with partial funding by the US DOE Office of Science via an award from The Isotope Development and Production for Research and Applications subprogram in the Office of Nuclear Physics.